\documentclass[11pt]{article}
\usepackage[margin=1.3in]{geometry}
\usepackage{amsfonts}
\usepackage{setspace}
\usepackage{hyperref}
\usepackage{enumitem}
\usepackage{verbatim}
\usepackage{amsmath}
\usepackage[round, sort]{natbib}

\usepackage{fancyhdr}

\title{Neo-Lorentzian Relativity and the Beginning of the Universe}
\author{Daniel Linford}

\begin{document}

\maketitle

\begin{abstract}
    Many physicists have thought that absolute time became otiose with the introduction of Special Relativity. William Lane Craig disagrees. Craig argues that although relativity is empirically adequate within a domain of application, relativity is literally false and should be supplanted by a Neo-Lorentzian alternative that allows for absolute time. Meanwhile, Craig and co-author James Sinclair have argued that physical cosmology supports the conclusion that physical reality began to exist at a finite time in the past. However, on their view, the beginning of physical reality requires the objective passage of absolute time, so that the beginning of physical reality stands or falls with Craig's Neo-Lorentzian metaphysics. Here, I raise doubts about whether, given Craig's NeoLorentzian metaphysics, physical cosmology could adequately support a beginning of physical reality within the finite past. Craig and Sinclair's conception of the beginning of the universe requires a past boundary to the universe. A past boundary to the universe cannot be directly observed and so must be inferred from the observed matter-energy distribution in conjunction with auxilary hypotheses drawn from a substantive physical theory. Craig's brand of Neo Lorentzianism has not been sufficiently well specified so as to infer either that there is a past boundary or that the boundary is located in the finite past. Consequently, Neo Lorentzianism implicitly introduces a form of skepticism that removes the ability that we might have otherwise had to infer a beginning of the universe. Furthermore, in analyzing traditional big bang models, I develop criteria that Neo-Lorentzians should deploy in thinking about the direction and duration of time in cosmological models generally. For my last task, I apply the same criteria to bounce cosmologies and show that Craig and Sinclair have been wrong to interpret bounce cosmologies as including a beginning of physical reality.
\end{abstract}

\hfill \break Forthcoming in the \emph{European Journal for Philosophy of Science}.

\hfill \break \textbf{Keywords:} time, Neo-Lorentzianism, beginning of the universe, Kalam argument, William Lane Craig, cosmology, relativity

\newpage

\doublespacing

\tableofcontents

\newpage

\begin{center}
    This paper is dedicated to my father, Paul Linford, who unfortunately passed away when this paper was accepted for publication. I love you, dad.
\end{center}

\section{Introduction}

Despite some disagreement over the last few decades, many physicists have maintained that absolute time was displaced by Einstein's relative time. Contrary to that tradition, William Lane Craig has long argued that relativity should be supplanted by an alternative, (supposedly) empirically indistinguishable, theory called \emph{Neo-Lorentzianism} (\cite{Craig_Rel:1990, Craig:1999, Craig:2001, craig_2001, Craig_balashov_reply, Craig:2008, Craig_Bergson}).\footnote{The case that relativity and the A-theory of time are not compatible has been presented in various places, but see \cite{Rietdijk:1966, Putnam:1967}; \cite[201, 303-304]{penrose:1989}; \cite{Petkov:2006, RomeroPerez:2014}. For work by physicists discussing Neo-Lorentzian theories, see \cite{Builder:1958, Prokhovnik:1963, Prokhovnik:1964a, Prokhovnik:1964b, Prokhovnik:1973, Prokhovnik:1986, Bell:1976, MacielTiomno:1985, MacielTiomno:1989a, MacielTiomno:1989b}. Balashov and Janssen (\citeyear{BalashovJanssen:2003}) have offered a masterful reply to Craig's Neo-Lorentzianism, to which Craig replies in his \citeyear{Craig_balashov_reply}. For a recent critical discussion of the view that absolute time is better accommodated by General Relativity than by Special Relativity because absolute time can be associated with cosmic time, as maintained by Craig and certain other A-theorists, see \cite{Read:2020}.} Elswhere, Craig, and his sometimes co-author James Sinclair, have argued that contemporary physical cosmology strongly supports the conclusion that the universe began to exist at a finite time in the past (\cite{Craig:1979, Craig:1992, Craig_Smith_1993A, Craig_1993_Criticism, CraigSinclair:2009, CraigSinclair:2012, CarrollCraig:2016}). Moreover, they have argued that \emph{beginning to exist} -- in their sense of the phrase -- is an irreducibly tensed notion, so that the universe could have begun to exist only if the A-theory of time -- that is, the view that there are objectively and irreducibly tensed facts -- is true (\cite[183-184]{CraigSinclair:2009}, \cite{Craig:2007}, \cite[337-338]{Craig_Rel:1990}); this conclusion is shared by many other philosophers and theologians, including William Godfrey-Smith (\citeyear{Godfrey-Smith:1977}), Bradley Monton (\citeyear[94]{Monton:2009}), David Oderberg (\citeyear[146]{Oderberg:2003}), Ryan Mullins (\citeyear[135-136, 143, 147]{Mullins:2016}; \citeyear[43]{Mullins:2011}), and Felipe Leon (\citeyear[62]{Leon:2019}). Other authors, e.g., \cite[11]{Reichenbach:1971}, hold that B-theory entails that nothing objectively begins or changes and so are implicitly committed to a view close to Craig and Sinclair's. For Craig, in light of the evidence supporting relativity, Neo-Lorentzianism is the only plausible choice for the A-theorist. If so, whether the universe began to exist at a finite time in the past stands or falls with Neo-Lorentzianism. I will refer to the conjunction of the A-theory and Neo-Lorentzianism as ANL. Here, I argue that when we take ANL much more seriously than many of Craig and Sinclair's interlocutors have in the past, we find that ANL invites a form of skepticism in tension with the view that present physical cosmology supports a beginning of the universe.

After unpacking and clarifying ANL, I consider two families of cosmological models. The first family of cosmological models are singular cosmologies (as explicated below); Craig and Sinclair have argued that singular cosmological models depict the universe as having had an \emph{ex nihilo} beginning at a finite time in the past. I argue that ANL invites skepticism in tension with Craig and Sinclair's treatment of singular cosmological models for two reasons. First, ANL severs the connection between our universe's matter-energy content and chronogeometric structure in such a way that we can no longer justifiably infer that our universe began to exist. Second, Craig and Sinclair have failed to adequately justify the adoption of a specific metric of absolute time. I offer an alternative metric for the duration of absolute time that is at least as good as (and possibly superior to) the one Craig and Sinclair favor. On the alternative metric, past time is infinite. For Craig and Sinclair, that the universe began to exist requires that past absolute time is finite. Without a way to adjudicate which of the two metrics -- if either -- corresponds to absolute time, we cannot infer whether past absolute time is finite. What these two epistemic challenges share is the realization that a beginning of the universe is not a directly observable feature of the universe; for that reason, the inference that the universe began to exist requires the conjunction of observation and a substantive physical theory. By displacing orthodox relativity with an underspecified alternative, Craig and Sinclair have effectively ripped away the ability they might otherwise have had to infer that the universe began. Along the way, we will see that there are additional reasons to think that while the conjunction of observation and a substantive physical theory may be necessary for inferring that the universe began, they may not be sufficient.
    
In thinking about singular cosmological models, I develop a set of criteria that friends of ANL can deploy to determine whether, according to any given cosmological model, physical reality had a beginning in the finite past. When I turn to a second family of cosmological models -- bounce cosmologies -- I utilize those tools in considering whether they represent physical reality as having begun to exist in the finite past. According to the orthodox interpretation of bounce cosmologies, a prior contracting universe birthed our expanding universe (e.g., \cite{Brandenberger2017, AshtekarSingh:2011, AgulloSingh:2017, ShtanovSahni:2003, Cai:2014jla, IjjasSteinhardt:2017, Ijjas_2018, Ijjas:2019pyf, Steinhardt:2002, steinhardt_2007, Corda:2011, Odintsov:2015, Oikonomou:2015}). Craig and Sinclair have argued for the reinterpretation of bounce cosmologies as the ex nihilo birth of two universes (a ``double bang'').\footnote{\cite{Huggett:2018} likewise argued that a variety of bounce cosmologies, particularly those developed utilizing loop quantum gravity, should be interpreted to involve a double bang and not a bounce. However, Huggett and Wüthrich's interpretation, as well as their argument for their interpretation, relies on the view that, in loop quantum gravity, time is explicable in terms of yet more fundamental physical entities. Craig and Sinclair's view involves a distinct metaphysics of time in which time is absolute and cannot be explicated in terms of yet more fundamental physical entities. Moreover, given that non-fundamentality of time in quantum gravity plausibly commits one to some variety of eternalism (see, e.g., \cite{Bihan:2020}), Huggett and Wüthrich's interpretation is plausibly incompatible with the universe having a beginning in Craig and Sinclair's sense.} In light of Craig's Neo-Lorentzianism, I extend arguments recently offered in \cite{Linford:2020a, Linford:2020b} against Craig and Sinclair's interpretation of bounce cosmologies. As I argue, just as ANL strips away the ability to infer that singular cosmologies include a beginning of the universe, so, too, ANL blocks the inference that bounce cosmologies depict physical reality as having begun to exist.

\section{The Beginning of the Universe, the Omphalos Objection, and ABIDO}

According to what J. Brian Pitts (\citeyear{Pitts:2008}) calls the \emph{metrical conception} of the beginning of the universe, the universe began to exist if space-time is finite to the past. Following Smith (\citeyear{Smith:1985}), Craig and Sinclair endorse the metrical conception:

\begin{quote}
    [...] we can say plausibly that time begins to exist if for any arbitrarily designated, non-zero, finite interval of time, there are only a finite number of isochronous intervals earlier than it; or, alternatively, time begins to exist if for some non-zero, finite temporal interval there is no isochronous interval earlier than it \cite[99]{CraigSinclair:2012}.
\end{quote}

However, Craig and Sinclair elsewhere indicate that, since beginning to exist is a tensed notion, the universe could only have begun to exist if the A-theory of time is true. Presentists, like Craig, maintain that future events do not yet exist, but come into being by becoming present, and past events become past by going out of existence. As Fay Dowker describes, ``The universe com[ing] into being [...] corresponds to the passage of time'' \cite[133]{Dowker:2020}. In contrast, B-theorists maintain that the distinction between the past, present, and future is perspectival and so future states of affairs do not not come into being by becoming present. For this reason, B-theory has sometimes been described as the view that space-time is objectively unchanging and eternal.\footnote{Of course, there are tenseless theories of change and perhaps one could utilize a tenseless theory of change to construct a tenseless conception of the beginning of time. Whatever one might think of the attempt to develop an account of the beginning of time utilizing a tenseless theory of change, my point is only that \emph{A-theorists} typically think objective change, and so the universe coming into being, requires the A-theory of time. 

Moreover, I take it that the A-theoretical conception of a beginning is particularly salient for theology. On many versions of A-theory, part of what it means for time to pass is (roughly) that each moment (somehow) produces a successor moment, with caveats for the possibility that time is continuous. And on the view that each moment must be produced -- either by another moment or by something else -- we can ask what could have produced the first segment of time, since the first segment of time could not have been produced by another time. Barring backwards causation -- which may not be possible on the A-theory anyway -- only some entity external to the temporal order could have produced the first segment of time.} As Craig and Sinclair write,

\begin{quote}
    On a B-Theory of time, the universe does not in fact come into being or become actual at the Big Bang; it just exists tenselessly as a four-dimensional space-time block that is finitely extended in the earlier than direction. If time is tenseless, then the universe never really comes into being, and therefore, the quest for a cause of its coming into being is misconceived \cite[183-184]{CraigSinclair:2009}.
\end{quote}

Elsewhere, Craig writes, ``The doctrine of creation involves an important metaphysical feature which is under-appreciated: it commits one to a tensed or A-theory of time. For if one adopts a tenseless or B-theory of time, then things do not literally come into existence. [...] The universe does not come into being on a B-theory of time, regardless of whether it has a finite or an infinite past relative to any time'' \cite[319]{Craig:2007}. In fact, Craig and Sinclair define \emph{beginning to exist} as a tensed fact. Quoting from the definition that they offer in their \cite[197]{CraigSinclair:2009}:

\begin{enumerate}
    \item  $x$ begins to exist at $t$ iff $x$ comes into being at $t$.
    \item $x$ comes into being at $t$ iff 
    \begin{enumerate}
        \item $x$ exists at $t$, and the actual world includes no state of affairs in which $x$ exists timelessly,
        \item $t$ is either the first time at which $x$ exists or is separated from any $t* < t$ at which $x$ existed by an interval during which $x$ does not exist, and
        \item $x$’s existing at $t$ is a tensed fact.\footnote{Craig \citeyear[99]{Craig:2002}; \citeyear[318]{Craig:2007} offer similar definitions; also see the discussion in \cite[337-338]{Craig_Rel:1990}.}
    \end{enumerate}
\end{enumerate}

Consequently, Craig and Sinclair implicitly accept that, in order to establish that the universe began to exist in Craig and Sinclair's sense, one must establish three claims. First, one must establish that the A-theory of time is true. Second, one must establish that we have good reason for identifying a past (closed or open) boundary to the universe, such that the universe did not exist before the boundary. And, third, one must establish that the span of absolute time between the past boundary and the present is finite. I do not claim that these three are sufficient for establishing that the universe began to exist, though Craig and Sinclair have not suggested any additional conditions, and I do not claim that there is no other conception of the beginning of the universe.\footnote{For a defense of the view that the universe beginning to exist does not require tensed facts, see \cite{Loke:2017}.} Instead, I claim that if these three desiderata cannot be satisfied, then the universe lacks a beginning in the sense that Craig and Sinclair have defended. In what follows, I examine the three criteria in order beginning with a somewhat lengthy review of ANL. Readers who are familiar with Neo-Lorentzianism may want to skip to section \ref{IDing_the_boundary} and refer back to section \ref{A-theory} only when necessary.

\subsection{\label{A-theory}The A-theory of Time and Neo-Lorentzianism}

Craig motivates his acceptance of A-theory on the grounds that the ``objectivity of tense and the reality of temporal becoming'' is a ``properly basic belief''. Craig goes on to write that ``belief in the reality of tense and temporal becoming enjoys such powerful positive epistemic status for us that not only can we be said to know that tense and temporal becoming are real, but also that this belief constitutes an intrinsic defeater-defeater
which overwhelms the objections brought against it'' \cite[138]{Craig:2000}. For Craig, that time passes is self-evident, immediate, and undeniable. Craig also maintains that only A-theory can successfully explain the directionality and anisotropy of time (\cite{Craig:1999}). Craig has argued that presentism is the only consistent version of A-theory and that presentism is incompatible with any plausible understanding of Special Relativity. Craig thinks that the only plausible way to save the A-theory of time, given the empirical adequacy of relativity within the relevant domain, is to accept a view I will call `Neo-Lorentzianism'.

To introduce Neo-Lorentzianism, and as a warmup exercise, let's first consider how Neo-Lorentzianism relates to orthodox Special Relativity. I will turn to General Relativity later, though I should note that, despite their interest in cosmology, neither Craig nor his co-authors have offered (or cited) a full-fledged Neo-Lorentzian alternative theory to General Relativity. 

In pre-relativistic physics, velocities are relative to reference frames. James Clerk Maxwell's discovery that his equations entail a unique value for the speed of light presented late nineteenth century physicists with two options. Either Maxwell's Equations do not hold in all reference frames or the measured value of the speed of light is the same in all reference frames. If the speed of light is the same in all reference frames, the Galilean transformations need to be replaced with Lorentz transformations (or something more complex). In turn, there are two ways to motivate the Lorentz transformations. First, in Special Relativity, one can accept the Lorentz transformations as primitives. Lengths, temporal durations, and simultaneity are relativized and no longer considered absolute; nonetheless, as Hermann Minkowski pointed out, the unification of space and time in a four-dimensional structure -- space-time -- can be considered absolute. Alternatively, following Lorentz, one can introduce a new set of forces that act on moving bodies to mimic the appearance of Special Relativity. Lorentzians agree that orthodox relativity is empirically adequate within its domain of application,\footnote{Some authors have argued that at least some forms of Neo-Lorentzianism can be empirically tested and so can be empirically distinguished from Special Relativity; see \cite{MacielTiomno:1985, MacielTiomno:1989a, MacielTiomno:1989b, Petkov:2006}.} but deny Minkowski's metaphysics. Their goal has been to restore absolute simultaneity, with the consequence that space and time are independent absolutes.\footnote{Neo-Lorentzianism does not, in itself, provide a conclusive answer to the debate between substantivalists and relationalists. As \cite[1]{Read:2020} explain, absolute time is ``defined as enshrining a universal present moment, objective temporal passage, and tensed facts''. Consequently, the existence of absolute space and absolute time do not entail that time is substantival, but does entail that spatial and temporal relations are observer independent.}

The trouble between presentism and Special Relativity arises because a moment of time, e.g., the present, as usually understood, consists of a collection of events, that is, space-time points, all of which are objectively simultaneous. Special Relativity appears to entail that there are no objectively simultaneous events. Therefore, if the objective present must consist of more than one event, Special Relativity appears to entail that there is no objective present and so to entail that presentism is false. Craig concludes that relativity should be supplanted by an alternative theory, i.e., Neo-Lorentzianism, that restores an absolute present.\footnote{To be sure, Neo-Lorentzianism is one among a number of different strategies that A-theorists might pursue in reply and A-theory might still be true if Neo-Lorentzianism is false; for recent surveys of possible replies, see (\cite{GilmoreCostaCalosi:2016, Baron:2018}). Nonetheless, Craig has argued against alternative A-theoretic strategies so that, on his view, the A-theorist has no plausible choice but to opt for Neo-Lorentzianism and so for ANL.}

For present purposes, let's set aside adjudicating the debate between A-theorists in order to focus on ANL. Neo-Lorentzians draw inspiration from Newton's \emph{Principia}. Newton maintained that absolute space should be distinguished from any physical measure of space; the change in spatial representation from one reference frame to another reflected differences in physical measures and not differences in space, as space is in itself. Likewise, Newton maintained a distinction between absolute time and physical measures of time \cite[6]{Newton:1974}; even though Newton did not consider the possibility that time would be measured differently from distinct reference frames, we can expect Newton to have said that differences in physical measures of time do not entail differences in how time is in itself. Neo-Lorentzians argue that, in order to maintain space and time as independent absolutes, we can re-deploy Newton's strategy with Einstein's coordinate transformations. Poincar{\'e} considered an important thought experiment that helps to illustrate this strategy. Poincar{\'e} imagined the interior of a sphere inhabited by creatures who carry rulers (\citeyear[55-57]{Poincare:2001b}). According to the laws of the world the creatures inhabit, as any object $O$ approaches the boundary of the sphere, the extension of $O$ in the direction of the sphere's radius shrinks to zero. As Poincar{\'e} points out, the creatures would mistakenly infer their world to be infinitely large. The lesson is that any attempt to measure the geometry of one's world will involve a decision about which effects are due to geometry and which effects are due to forces acting on one's measuring instruments. By carefully selecting a set of forces, measurements that appear to confirm Minkowski space-time can be rendered consistent with Newtonian absolute space and time.

Lorentz considered the possibility that forces conspire to obscure our world's true chronogeometry in such a way that the speed of light appears the same from every reference frame. For example, in order to maintain equilibrium with the surrounding electromagnetic field, a moving spherical distribution of charge would need to foreshorten in precisely the way that Einstein predicted objects foreshorten when they move close to the speed of light. Neo-Lorentzians employ different mechanical models that they think might explain relativistic phenomena. For example, Simon Prokovnik (\citeyear[79-89]{Prokhovnik:1986}), an important proponent of Neo-Lorentzianism (\citeyear{Prokhovnik:1963, Prokhovnik:1964a, Prokhovnik:1964b, Prokhovnik:1973, Prokhovnik:1986}),  points out that if Newtonian gravitation is altered to include the fact that gravitational influences propagate at the speed of light (as previously suggested in, e.g., \cite{bastin:1960}), then moving bodies would need to foreshorten to maintain equilibrium with the gravitational field. Prokhovnik argues that the other predictions made by Special Relativity -- including the appearance of the relativity of simultaneity -- can similarly be recovered from gravitational phenomena. More recent work has shown that while one can always redescribe relativistic effects as the result of a universal force, the forces that one must postulate have fairly exotic (arguably, implausible) properties (\cite{WeatherallManchak:2014}).\footnote{An anonymous reviewer has brought to my attention that teleparallel gravity may offer another approach to redescribing relativistic effects as the result of forces, e.g., \cite{Knox:2011}.}

Although Craig consulted Prokovnik while writing his book length defense of ANL (\citeyear{Craig:2001}) and cites Prokovnik throughout, Craig sets aside whether Prokhovnik's explanation in terms of gravitation is correct \cite[181]{Craig:2001}. Following \cite{MacielTiomno:1989a}, Craig thinks of Neo-Lorentzianism as a family of theories: 

\begin{quote}
    A theory may be classified as [Neo-]Lorentzian just in case it affirms (i) physical objects are $n$-dimensional spatial objects which endure through time, (ii) the round trip vacuum propagation of light is isotropic in a preferred (absolute) reference frame $R_o$ (with speed $c=1$) and independent of the velocity of the source, and (iii) lengths contract and time rates dilate in the customary special relativistic way only for systems in motion with respect to $R_o$ \cite[178]{Craig:2001}.
\end{quote}

On Craig's brand of Neo-Lorentzianism, physical objects are three dimensional entities that endure through time. The present can be approximated as a three-dimensional hypersurface that changes as time passes.\footnote{There are some additional nuances in Craig's view, since Craig denies the existence of instants.  We might have expected Craig to endorse the existence of instants given his presentism, but Craig has long maintained that instants do not exist. For example, Craig argues that ``only intervals of time are real or present and that the present interval (of arbitrarily designated length) may be such that there is no such time as `the present' \emph{simpliciter}; it is always `the present hour', `the present second', etc. The process of division is potentially infinite and never arrives at instants'' \cite[260]{Craig_1993_Criticism}; also see \cite[179-180]{Craig:2000_extent}. Craig maintains that time is gunky, i.e., that every interval of time has proper sub-intervals, and he maintains that time cannot be decomposed into instants one of which is the present. I confess that I find the conjunction of presentism and gunky time difficult to understand. For example, if presentism is the thesis that the only things that exist simpliciter are present and that there is no present simpliciter -- as Craig's gunky time seems to entail -- does nothing exist simpliciter?}

\subsection{\label{IDing_the_boundary}Identifying the Universe's Past Boundary}

Having reviewed the first criterion for establishing that the universe began -- viz., ANL -- let's turn to the second criterion. The second criterion involves a past boundary for the universe beyond which the universe's past cannot be extended. In pre-relativistic physics, the universe's contents cannot be used to determine when (or whether) time began. Consider, for example, an acorn. Suppose that God created the acorn \emph{ex nihilo} last Thursday. In that case, we'd have no indication that the acorn did not exist before last Thursday because the acorn's existence instead suggests the existence of a pre-existing tree that gave rise to the acorn. Likewise, the entire universe could have been created last Thursday, even though we have memories of prior times. This suggests that a beginning of the universe would make no difference to the universe's matter-energy content and so would be empirically undetectable. These considerations give rise to a general skeptical problem: 

\begin{enumerate}
    \item No matter when the universe began, the universe's contents suggest a prior history.
    \item If no matter when the universe began, the universe's contents suggest a prior history, then we cannot empirically determine when the universe began.
    \item Therefore, we cannot empirically determine when the universe began.
\end{enumerate}

In \citeyear{Gosse:1857}, Philip Gosse voiced a related argument in his book \emph{Omphalos: An Attempt to Untie the Geological Knot}; therefore, call this worry the \emph{Omphalos Objection}.\footnote{For Gosse, the view that we cannot empirically determine whether -- or when -- the universe began was part of a strategy to render a literal reading of the Bible compatible with evidence from geology that the Earth is older than the Biblical narrative appears to indicate. Nonetheless, Gosse's strategy has never enjoyed popularity -- as Russell wrote, ``nobody can believe it'' (\citeyear[68]{Russell:1961}) -- and, in any case, is straightforwwardly incompatible with any attempt to empirically determine the age of either the Earth or the universe.}

Perhaps one could object that even if we do not know when the universe began, one may still be able to infer that the universe began. For example, perhaps one can produce a rationalist argument that any temporal (or causal) series must have a finite length. However, this would be an admission that, at least for friends of ANL, an \emph{empirical} case for a beginning of the universe does not succeed and so would be a concession to the arguments that I offer in this paper. I have difficulty conceiving of a situation in which we have \emph{empirical} grounds for inferring that the universe began to exist but we do not have empirical grounds for inferring \emph{when} the universe began to exist. In the case of the Earth, we have evidence that the Earth began to exist precisely because we have evidence for when the Earth began to exist. If we accept Craig and Sinclair's analysis of \emph{beginning to exist}, an empirical case for the beginning of the universe would presumably share this feature with the empirical case for the Earth's beginning.

Relativistic physics appears to resolve the Omphalos Objection for two reasons. First, among relativists, a common working assumption -- that is, the assumption that space-time is maximally extended -- prohibits truncated space-times (e.g., \cite[32]{Earman:1995}). As Earman has argued, allowing space-time to be prematurely truncated reintroduces a version of the Omphalos Objection and results in a skeptical catastrophe \cite[119-122]{Earman:1977}. Second, although we cannot observe the chronogeometric structure of space-time directly, we can observe the distribution of our universe's matter-energy content. In turn, General Relativity ties the matter-energy distribution to the geometry of space-time in such a way that, by observing the matter-energy content, we may be able to infer that our space-time has a past open boundary beyond which space-time cannot be extended.\footnote{A similar point was previously raised in, e.g., (\cite{Weingard:1979}).} Arguably, this is precisely the case in classical models of the Big Bang. For example, in Friedmann-Lema{\^i}tre-Robertson-Walker (FLRW) models, the scale factor is zero for some value of the cosmic time, thereby resulting in a corresponding divergence in the energy density. The divergence in the energy density corresponds to a divergence in the Ricci scalar. The result is a space-time singularity,\footnote{A divergence in the various curvature parameters is neither necessary nor sufficient for space-time to be singular (\cite{Earman:1995, Curiel:1999, Curiel:2021, Joshi:2014}). For example, so-called conical singularities are well-known examples of singularities without associated curvature pathology. However, the singularity that represents the ``Big Bang'' in cosmologically relevant space-times is associated with a divergence in the Ricci scalar and the energy density.} i.e., an open boundary beyond which space-time cannot be extended. Thus, so long as we realistically construe the relationship between the universe's matter-energy content and the geometry of space-time, and assume that space-time is maximally extended, then there is hope that a survey of the universe's matter-energy content would indicate a past open boundary to space-time. (I say `hope' because, as we will see below, there is reason to doubt that this hope can be fulfilled.)

My description of Craig and Sinclair's views would be incomplete without remarking on the fact that most physicists view divergences in physical theories with suspicion.\footnote{While the predictions of a theory within a specific domain may provide some inductive evidence that the theory will apply to neighboring domains, no one should have confidence that the theory will apply to domains that are arbitrarily distant. Consider approaching a point $p$ where the energy density diverges. As one approaches $p$, one encounters arbitrarily large energy densities and so one inevitably encounters energy densities which surpass the domain of applicability of General Relativity before one reaches $p$. For that reason, ceteris paribus, we should doubt the predictions made by General Relativity within the vicinity of curvature singularities.} For that reason, most physicists think we should not draw conclusions about the beginning of the universe from singular cosmological models. Craig and Sinclair maintain instead that singular cosmological models \emph{do} provide strong evidence for a beginning of the universe because they argue that there will be features in a future quantum gravity theory that correspond to the cosmological singularities in FLRW models. As they write, ``There may be no such things as singularities per se in a future quantum gravity formalism, but the phenomena that [General Relativity] incompletely strives to describe must nonetheless be handled by the refined formalism, if that formalism has the ambition of describing our universe'' \cite[106]{CraigSinclair:2012}. I don't find this reply convincing. However, whether or not we should look upon divergences in physical theories as suspicious has been discussed at length elsewhere and I set the issue aside for the purposes of this paper.

For the sake of clarity, the argument that I am considering here -- which rules out extendable space-times in order to avoid the Omphalos Objection -- is distinct from another set of arguments that have been offered by John Earman and criticized by J.B. Manchak. For example, Earman (\citeyear[32-33]{Earman:1995}) has argued that extendable space-times can be thrown away by invoking metaphysical principles such as the principle of sufficient reason. Manchak rejects Earman's metaphysical argument. Moreover, Manchak (\citeyear{Manchak:2011}) has argued that we cannot determine, on empirical grounds, whether the space-time we inhabit is maximally extended. In any case, neither argument is the argument that I am considering here; the argument under consideration does not rule out extendable space-times for metaphysical or empirical reasons. Instead, the argument that I am considering rules out extendable space-times on epistemic grounds; that is, if we allow for extendable space-times, then a skeptical catastrophe results. For example, if space-time can be arbitrarily truncated, then all of our memories -- including our memories of whatever scientific experiments or observations we take to support our best theory of space-time -- could have been created ex nihilo last Thursday. As Bertrand Russell wrote, ``We may all have come into existence five minutes ago, provided with ready made memories, with holes in our socks and hair that needed cutting'' (\citeyear[68]{Russell:1961}).\footnote{An anonymous reviewer suggests that we may have reasons for trusting our memory that differ from the reasons we have for trusting other sorts of records of the past. For example, perhaps we have some reason to think that God would ensure our memory of the past is veridical. If so, we may have reason to avoid truncating the past during any period that any person remembers without invoking a general prohibition on truncated space-times, so that we may be able to avoid a skeptical catastrophe without invoking a prohibition on truncated space-times. Nonetheless, we ordinarily take our empirical access to the past to be based on more than memories. Importantly, successful cosmological science requires ampliative inferences to time periods no mere mortal could remember, e.g., the early universe. Perhaps we could once more invoke God to secure the veridicality of our records or of other ampliative inferences we might make to the past, but this begins to look like yet another prohibition on truncated space-times, albeit a prohibition invoking theological premises. Craig and Sinclair are unlikely to pursue this route, for they utilize cosmological science in defense of God's existence, and naturalists are unlikely to find a theological defense of a truncation prohibition principle convincing.} And, if we recognize this as a genuine possibility, we would have undermined whatever support we take ourselves to have for our best theory of space-time in the first place. Thus, if we accept as a live possibility that space-time could be arbitrarily truncated, we undercut the evidence that we have for our best theory of space-time. And if the evidence we have for our best theory of space-time is undercut, we lose whatever support we would otherwise have had for our best theory of space-time. To avoid the skeptical catastrophe, we should rule out extendable space-times.\footnote{Parallel arguments have been used for including the Past Hypothesis as a fundamental physical principle \cite[116]{albert_2000} and to rule out cosmologies in which Boltzmann brains dominate (\cite{Carroll:2021}).}

I have said that General Relativity may provide \emph{hope} for resolving the skeptical catastrophe posed by postulating a beginning of the universe in the finite past. However, these hopes are problematized -- and possibly dashed -- by a series of recent results formulated by Manchak. As I've said, the skeptical catastrophe is supposedly avoided by throwing away space-times that are not maximally extended. However, that space-time $(M, g_{\mu\nu})$, where $M$ is a pseudo-Riemannian manifold with metric tensor $g_{\mu\nu}$, fails to be maximally extended is likely not the only condition that would render $(M, g_{\mu\nu})$ physically unreasonable. If the A-theory of time is true, then the A-theory of time is likely necessarily true, or at least true at all of the physically possible worlds. And the A-theory of time -- or at least the version of A-theory that Craig has defended -- appears to require that space-time is (for example) globally hyperbolic. Suppose, then, that all physically reasonable space-times are globally hyperbolic. In that case, let's say that $(M, g_{\mu\nu})$ is maximally GH-extended just in case $(M, g_{\mu\nu})$ has no globally hyperbolic extensions. 

Following Manchak (\citeyear{Manchak:2021}), consider that Misner space-time is not globally hyperbolic and so, if only globally hyperbolic space-times are physically reasonable, Misner space-time would not be considered physically reasonable. But there is a region of Misner space-time that is hyperbolic. Consider then a model of space-time consisting only of this region excised from Misner space-time; this new space-time, though not maximally extended, is globally hyperbolic. Moreover, as Manchak has argued, this new space-time has no globally hyperbolic extensions. Consequently, the new space-time is maximally GH-extended even though the new space-time is not maximally extended. So: if we say that `maximally GH-extended space-time' and `physically reasonable space-time' are co-extensive, we will have picked out a different collection of space-times than if we say that `maximally extended space-time' and `physically reasonable space-time' are co-extensive. Furthermore, a similar argument can be used to show that if we select some proper subclass of maximally GH-extended space-times satisfying (say) property $P$, then there will generally be space-times that are maximally $P$-extended but not maximally GH-extended.

If, for example, `maximally GH-extended space-time' and `physically reasonable space-time' are co-extensive, then the globally hyperbolic region we excised from Misner space-time has a boundary. Therefore, upon discovering local evidence that one inhabits a Misner space-time (for example) whether one would then have grounds for inferring that one's universe has a boundary will depend upon how physical reasonableness should be understood.

Obviously, this creates a conceptual difficulty for how we should understand the relationship between the extendability of space-time and the resolution of the Omphalos Objection. Work is on-going and there is, as yet, no consensus as to what features make a space-time physically reasonable. The relevant question for Craig and Sinclair is whether a new set of criteria for physical reasonableness can be successfully developed and agreed upon that would allow one to identify a past boundary to space-time by examining the mass-energy-momentum distribution within space-time. For example, if global hyperbolicity does turn out to be the relevant ``mark of physical reasonableness'' and one could show, from the matter-energy distribution, that our space-time cannot be GH-extended beyond some boundary in the finite past, then Craig and Sinclair would have one of the necessary ingredients for a successful empirical case for a beginning of the universe. As of yet, such a result has failed to materialize.

While Manchak has produced a series of results problematizing the notion of physical reasonableness, Manchak and others have simultaneously produced a series of results casting doubt on our ability to determine the global structure of space-time. Roughly, space-time $(M, g_{\mu\nu})$ is \emph{observationally indistinguishable} from some numerically distinct space-time $(M', g'_{\mu\nu})$ just in case any observer at any arbitrarily chosen point in $(M, g_{\mu\nu})$ cannot determine, from \emph{any} of the data in their past light cone, which of the two space-times they inhabit \cite[412]{Manchak:2011}. As the Manchak-Malament theorem has established (\cite{Malament:1977, Manchak:2009}; also see \cite{ Manchak:2021, Norton:2011, Beisbart:2009, Butterfield:2014}), the members of a fairly broad class of space-times are observationally indistinguishable from numerically distinct space-times. 

To pose an epistemological challenge, one requires a significantly weaker condition than observational indistinguishability, at least as previously defined. Observational indistinguishability imposes a condition on \emph{any} arbitrarily chosen observer and consequently for any space-time point. For us to be unable to empirically distinguish our space-time from a space-time with quite different global properties requires only a result on data that we can gather from within our cosmological horizon and within some reasonable proper time. And \emph{this} entails that our space-time is indistinguishable -- as far as we are concerned -- from an even larger class of space-times. In order to distinguish this form of observational indistinguishability from those catalogued in \cite{Malament:1977}, I will denote it \emph{super weak observational indistinguishability}.

What is relevant for my purposes in this paper is that singular FLRW space-times -- that is, FLRW space-times such that all time-like or null curves are inextendable beyond some boundary located in the finite past -- are super weakly observationally indistinguishable from space-times containing at least one time-like or null curve that is not bounded to the past. For example, the Borde-Guth-Vilenkin (BVG) theorem (\cite{Borde:2003}) has sometimes been interpreted to show that the universe has a past singular boundary. However, the BVG theorem actually shows that if the average value of a specific generalization of the Hubble parameter along a time-like or null geodesic congruence is positive, then that congruence must be incomplete to the past. The resulting congruence can be isometrically embedded either into a space-time with or without a global past boundary. That is, supposing that we inhabited such a congruence, we might see a boundary to our past, even though other observers in the same space-time would have an infinite and unbounded past.

So, there is reason to doubt, first, that space-time extendability's relationship to a beginning of the universe has been successfully characterized and, second, that, even given a correct characterization, we could have epistemic access to sufficient data to determine important global properties of the space-time we inhabit, including whether our space-time has a global boundary in the finite past. Nonetheless, we also have reason to think that if a beginning of the universe in the finite past could be empirically established with the resources of relativistic cosmology, then empirically establishing a beginning of the universe in the finite past would draw upon results concerning maximal extendability, or whatever the appropriate notion of extendability turns out to be. Any such result depends upon an inference from the distribution of the matter-energy contents of space-time to the chronogeometry of space-time. That is, results about extendability, in whatever the appropriate sense turns out to be, are likely to be necessary, but not sufficient, for using the resources of relativistic cosmology to establish a beginning of the universe in the finite past. And as I argue in the next section, insofar as there is tension between ANL and our being able to empirically infer the actual chronogeometry of our universe, then there is tension between ANL and the use of extendability results in resolving the Omphalos Objection.

\subsubsection{Tension between ANL and the resolution of the Omphalos Objection}

I've argued that relativistic cosmology may help in resolving the Omphalos Objection. So long as we realistically construe the relationship between the universe's matter-energy content and space-time curvature, and assume that space-time is maximally extended (in whatever sense turns out to be appropriate), then there is hope that a survey of the universe's matter-energy content would indicate a past boundary to space-time. We've also seen that there may be some reason to think that this hope has been put in doubt. But insofar as relativistic cosmology can be utilized to empirically determine whether our universe began to exist in the finite past, such a determination can be made only by utilizing the observed matter-energy content and the relationship relativity provides us between that content and chronogeometric structure. Friends of ANL do not realistically construe the relationship between the universe's matter-energy content and space-time curvature. They maintain that our epistemic situation with respect to absolute space and absolute time resembles the epistemic situation of  Poincar{\'e}'s creatures in that our world's true chronogeometry is hidden from us. If our world's true chronogeometry is hidden from us, then there is the possibility that the universe only appears to have a past boundary. What  we  need  to  know  is  not  whether our  universe  merely \emph{appears}  to  have  a past boundary when measured by physically embodied observers, but enough about physical reality’s relationship to absolute time to know that the universe \emph{really does} have a past boundary.

To put the point another way, consider a classic argument for space-time conventionalism that builds on Poincar{\'e}'s previously mentioned thought experiment. Poincar{\'e}, Reichenbach, Gr{\"u}nbaum, and other space-time conventionalists maintain that significant chronogeometric features, such as simultaneity, are conventional. For Reichenbach, space-time conventionalism can be supported by noting that any empirical determination of chronogeometry will need to adopt a convention as to which effects are due to chronogeometry and which effects are due to forces that universally act on the objects populating space-time \cite[30-34, 118-119]{Reichenbach:1958}. Philosophy of science has come a long way since the logical positivists. For that reason, although space-time conventionalism still retains proponents, philosophers will tend to see Reichenbach as having moved too quickly; the fact that we cannot empirically determine which effects are due to forces and which effects are due to chronogometry -- if it is a fact -- does not suffice for the conclusion that there is no fact of the matter concerning which effects are due to forces and which effects are due to chronogeometry.

Those who take a realistic interpretation of relativity set the universal forces to zero. In contrast, Craig and Sinclair believe we have independent reason to maintain absolute time and, for that reason, Craig and Sinclair endorse the existence of non-zero universal forces whose effect is to make our world appear as though relativity were true. The trouble is that the past boundary postulated by relativistic cosmology is just yet more chronogeometric structure; what Craig and Sinclair need to provide us is a way to infer, from empirical observations, enough about absolute time so that we can infer that absolute time has a beginning in the finite past. But this doesn't seem possible. As John Norton writes, universal forces are ``entities protected from evidential scrutiny by careful contrivance'' (\cite{Norton:2020}). As in Poincar{\'e} and Reichenbach's original argument, without being able to determine the universal forces operating on bodies, we are likewise left without a way to determine the fundamental chronogeometry.

There are various routes Craig and Sinclair might pursue to resolve the tension between inferring that absolute time began in the finite past and their view that the actual chronogeometry is protected from evidential scrutiny. For example, some A-theorists have maintained that space and time have the structure postulated by relativity conjoined with some additional structure that they attribute to absolute time; call this view $GR^+$. Because $GR^+$ is logically stronger than General Relativity, any space-time that is singular with respect to General Relativity is likewise singular with respect to $GR^+$. (Likewise, any space-time that is inextendable with respect to General Relativity for some other reason -- for example, any space-time that has no GH-extension -- will likewise be inextendable with respect to $GR^+$.) Unfortunately, Craig and Sinclair do not maintain $GR^+$. Friends of $GR^+$ maintain that space-time fundamentally has the structure postulated by General Relativity and therefore claim that General Relativity has some identifiable significance for fundamental metaphysics. In contrast, Craig and Sinclair maintain that space and time do not have the structure postulated by General Relativity; instead, space and time have some altogether different structure, appearing to satisfy General Relativity only because some collection of universal forces -- left unspecified -- distort all of our measuring instruments in just the right way so as to render General Relativity empirically adequate. Whatever the underlying space-time may be, Craig and Sinclair can always postulate some set of universal forces that appropriately distorts chronogeometric measurements. (A similar point was previously made in, e.g., \cite[15]{Read:2020}.) In some sense, this was the point originally made by the space-time conventionalists, that is, our measurements can be rendered consistent with any chronogeometry whatsoever so long as one is sufficiently creative with the forces that one postulates. ANL thereby severs the connection from physical measures of space and time to the true chronogeometry. 

On the resulting instrumentalist interpretation of length contraction, time dilation, and the relativity of simultaneity, rulers and clocks provide systematically spurious results due to the influence of universal forces. For that reason, rulers and clocks are no help in determining our world's true chronogeometry. Furthermore, length contraction, time dilation, the relativity of simultaneity, and other consequences of Special Relativity are ordinarily thought to be consequences, at least in part, of the metric $g_{\mu\nu}$; after all, one way to \emph{derive} the various consequences of Special Relativity begins with $g_{\mu\nu}$. If length contraction, time dilation, the relativity of simultaneity, and other relativistic effects are to be treated instrumentally because they are subject to the influence of universal forces, we should say that $g_{\mu\nu}$ is merely an apparent metric that affords empirically adequate predictions and so does not reflect the true metric of the underlying chronogeometry. That is, $g_{\mu\nu}$ is afforded an instrumental interpretation in Special Relativity. But if $g_{\mu\nu}$ is afforded an instrumental interpretation in Special Relativity, $g_{\mu\nu}$ should equally be afforded an instrumental interpretation in General Relativity. Likewise, the motion of test particles cannot help Craig and Sinclair in determining the true chronogeometric structure because the motion of test particles will be subject to the same universal forces distorting our measuring instruments.

Consider that, in classical electrodynamics, the electric and magnetic fields are invisible. In order to ``see'' the electric and magnetic fields, we need to utilize the fact that test bodies, e.g., iron filings and small charges, couple to the electric and magnetic fields via the Lorentz force law. In turn, the dynamics of the electric and magnetic fields are described by Maxwell's equations. Analogously, on a realist interpretation of General Relativity, test masses reveal chronogeometric structure because the trajectories of test masses are described by the geodesic equation. That is, the way in which test masses couple to space-time is an important auxiliary hypothesis for inferring chronogeometry from observational data.\footnote{\label{aux-hyp}Perhaps one can object that, in General Relativity, other mathematical relationships can describe the way that matter-energy couples to chronogeometry than the geodesic equation, for example, by the source term in the Einstein Field Equation or the Raychaudhuri equation. And, arguably, the Raychaudhuri equation is more important for inferring singular behavior from the matter-energy distribution. But analogous conclusions follow; whatever auxiliary hypothesis one uses, so long as the auxiliary hypothesis follows from General Relativity, an inference to the actual, and not merely apparent, chronogeometry requires a realistic construal of General Relativity.} As Misner, Thorne, and Wheeler famously quipped, ``Space tells matter how to move. Matter tells space how to curve'' (\citeyear[5]{MisnerThorneWheeler:1973}). This is the insight that, on a realistic construal of General Relativity, allows us to infer invisible chronogeometric structure from the visible distribution of matter. On a realistic construal, the Einstein Field Equations (together with the geodesic equation) express a relationship between the matter-energy distribution and space-time curvature, i.e.,

\begin{equation}
  G_{\mu \nu} = 8\pi G T_{\mu \nu}  
\end{equation}

Here, $G_{\mu \nu}$ is the Einstein tensor and is comprised by the Ricci curvature tensor and the Ricci curvature scalar while $T_{\mu \nu}$ is the stress-energy tensor, expressing the matter-energy distribution. In order to compute trajectories of test bodies, one can utilize the geodesic equation:

\begin{equation}
    \frac{d^2 x^{\mu}}{ds^2} + \Gamma^{\mu}_{\hphantom{\mu}\alpha\beta}  \frac{d x^{\alpha}}{ds} \frac{d x^{\beta}}{ds} = 0
\end{equation}

$x^\mu$ is the set of coordinates specifying a trajectory parametrized by $s$ and $\Gamma^{\mu}_{\hphantom{\mu}\alpha\beta}$ is the Christoffel symbol computed from the relevant metric as obtained from the Einstein Field Equations.\footnote{The exact logical relationship between the Einstein Field Equations and the geodesic equation has been the matter of some dispute. For example, the Geroch-Jang theorem, as well as various related results, show that, at least for space-times and matter satisfying a small set of realistic conditions, the motion of small massive bodies (e.g., test masses) satisfies the geodesic equation. See \cite{GerochJang:1975, EhlersGeroch:2004, Brown:2005, Weatherall:2011, Weatherall:2019, Malament:2012}. For my purposes, the point is that General Relativity provides us with a set of mathematical principles, whatever their interrelationship might be, which, when realistically construed, allow us to infer chronogeometric structure from the mass-energy distribution.} On the implementation of universal forces described by Michael Friedman (\citeyear[298]{Friedman:1983}), the geodesic equation is modified to accommodate the universal force $F^{\mu}$ and an associated metric compatible with a connection given by $\tilde{\Gamma}^{\mu}_{\hphantom{\mu}\alpha\beta}$:

\begin{equation}
    \frac{d^2 x^{\mu}}{ds^2} + \tilde{\Gamma}^{\mu}_{\hphantom{\mu}\alpha\beta}  \frac{d x^{\alpha}}{ds} \frac{d x^{\beta}}{ds} = F^{\mu}
\end{equation}

In turn, $F^{\mu}$ can be calculated in terms of the original and modified connections:

\begin{equation}
    F^{\mu} = (\tilde{\Gamma}^{\mu}_{\hphantom{\mu}\alpha\beta} - \Gamma^{\mu}_{\hphantom{\mu}\alpha\beta})\frac{d x^{\alpha}}{ds} \frac{d x^{\beta}}{ds}
\end{equation}

Since the modified geodesic equation is trivially equivalent to the original geodesic equation, the modified geodesic equation will result in the same trajectories as the original geodesic equation. Given the tremendous range of freedom in the specification of $F^{\mu}$, we have a corresponding freedom in how we specify $\tilde{\Gamma}^{\mu}_{\hphantom{\mu}\alpha\beta}$ and therefore in how we specify the metric. Furthermore, Friedman's proposal represents merely one way, out of a myriad of possibilities, for specifying universal forces. More radical proposals might replace the geodesic equation altogether.

An instrumental interpretation accepts the observable matter-energy distribution and accepts the Einstein Field Equation and geodesic equation, but only as useful calculational devices. Instrumentalists deny that the Einstein Field Equation and geodesic equation have ontological import for inferring unobservable chronogeometric structure. Thereby, instrumentalists sever the inference from the matter-energy distribution to the \emph{real} chronogeometry.\footnote{Craig provides us with another reason for thinking that the fundamental chronogeometric structure is decoupled from the motion of test masses. Though Craig does not endorse Prokovnik's views about gravity, Craig does maintain an instrumentalist interpretation of the relationship between the distribution of matter-energy and chronogeometric structure. For example, Craig writes that the ``geometrization of gravitation'' is only ``a heuristic device'' \cite[189]{Craig:2001}. Elsewhere, Craig and Sinclair explicitly deny the ``view that gravity \emph{just is} the curvature of an objectively real space-time'' (emphasis is Craig and Sinclair's) and instead argue that gravity is a force that operates between bodies situated in space \cite[104]{CraigSinclair:2012}. If, as they write, the ``geometrization of gravity'' is a mere ``heuristic device'' and gravity is instead a force operating between bodies, then the dynamics of the matter-energy distribution should ultimately be explained in terms of a force instead of the coupling of the matter-energy distribution to space-time curvature demanded by the Einstein Field Equation.

I'm not sure what sort of forces Craig and Sinclair would put in place of space-time curvature; they never offer a fully worked out and mathematically precise alternative to General Relativity. \cite[189]{Craig:2001} cites \cite[vii]{Weinberg:1972}, but Weinberg alternately states that his focus on geometry is a pedagogical strategy instead of a denial that gravity is the curvature of space-time (\citeyear[viii]{Weinberg:1972}) and that he is otherwise ambivalent concerning the metaphysical upshot of General Relativity: ``The important thing is to be able to make predictions on the astronomers' photographic plates, frequencies of spectral lines, and so on, and it simply doesn't matter whether we ascribe these predictions to the effects of gravitational fields on the motion of planets and photons or to a curvature of space and time.'' In other words, Weinberg's attitude -- at least as of 1972 -- was that, instead of trying to determine the metaphysics of space-time, we should ``shut up and calculate''. This is obviously not an attitude that friends of ANL can adopt. Moreover, Weinberg is no friend of Craig's approach to relativity. Weinberg's anti-metaphysical interpretation of relativity is likely the result of a wholesale anti-metaphysical attitude that would reject appeals to absolute time and absolute space. Elsewhere, Weinberg has argued that we should ``score'' physical theories against whether they satisfy Lorentz invariance (\citeyear[85]{Weinberg:2003}) -- whereas ANL only \emph{appears}, but does not actually, satisfy Lorentz invariance -- and that we should think Einstein was rightly victorious in his debate with Lorentz (\citeyear[68, 85]{Weinberg:2003}).}

Singular space-times are typically identified in virtue of geodesic incompleteness. Realists utilize the Einstein Field Equations, the geodesic equation, and the observed matter-energy distribution (or some other auxiliary hypothesis, i.e., see footnote \ref{aux-hyp}) to infer that space-time is geodesically incomplete to the past. Since geodesic incompleteness is a bit of chronogeometric structure to which the insrumentalist is not metaphysically committed, by endorsing an instrumental interpretation of the Einstein Field Equations and related mathematical relationships, Craig and Sinclair cannot justifiably use the Einstein Field Equations, the geodesic equation, or other auxiliary hypotheses from General Relativity, and the observed matter-energy distribution to infer that space-time really is geodesically incomplete to the past. 

In addition to geodesic incompleteness, singular FLRW space-times are characterized by a divergent Ricci scalar curvature. The realist can utilize the Einstein Field Equations and the matter-energy distribution to infer that the curvature was arbitrarily large in the past. The instrumentalist affords the various curvature parameters -- and so curvature pathology -- an instrumental interpretation and so is not committed to the ontological reality of curvature pathology. For that reason, Craig and Sinclair cannot infer that the Ricci scalar (or other curvature parameters) were arbitrarily large in the past.\footnote{Perhaps I've moved too quickly here. As an anonymous reviewer points out, while Craig and Sinclair cannot infer that the Ricci scalar, or other curvature parameters, qua curvature of space-time, was arbitrarily large in the past, they may be able to infer that the Ricci scalar, or other curvature parameters, construed as some mixture of geometry and universal forces, was arbitrarily large. 

Two comments can be made in reply. First, any argument from curvature pathology to singular behavior is weak because curvature pathology does not ensure a space-time singularity. Part of what matters for a boundary to space-time is that there is some ``location'' beyond which paths cannot be reasonably extended and yet ``[...] no species of curvature pathology we know how to define is either necessary or sufficient for the existence of incomplete paths'' (\cite{Curiel:2021}). For this reason, space-time singularities are now typically understood in terms of geodesic incompleteness and not in terms of curvature pathology. Second, a physical field can exhibit singular behavior without a corresponding boundary to time. For example, in classical electrodynamics, electric charges are singularities in the electric field. Classical electrodynamics is well-defined on Minkowski space-time, for which there is no past boundary. If we understand $g_{\mu\nu}$ as a physical field defined on a background absolute space and time, then, instead of attributing curvature pathology to an objectively real temporal boundary, curvature pathology can be attributed to a divergence in a physical field. In that case, we come to the analogy that I construct between ANL and the theory considered by Feynman, Pitts, and Schieve.}

There is an intriguing analogy between ANL and a theory considered by Richard Feynman (\citeyear{Feynman:2003}) and by J. Brian Pitts and W. C. Schieve (\citeyear{PittsSchieve:2003, PittsSchieve:2004, PittsSchieve:2007}; also see \cite{Pitts:2019}).\footnote{\cite{PittsSchieve:2007} and \cite{Pitts:2019} consider another similar theory that is, in principle, empirically distinguishable from standard General Relativity. The theory approximates standard General Relativity arbitrarily well given a sufficiently small graviton mass.} According to the theory Feynman, Pitts, and Schieve consider, physicists have been wrong to think of the metric $g_{\mu\nu}$ appearing in General Relativity as a description of space-time; instead, Pitts and Schieves consider the possibility that like, e.g., the electromagnetic field, $g_{\mu\nu}$ is a gravitational field (i.e., the field of a spin-2 boson) defined on a background Minkowski (flat) space-time equipped with a metric $\eta_{\mu\nu}$. Therefore, although Craig and Sinclair's brand of ANL maintains a background absolute space and absolute time, and Minkowski space-time differs from absolute space and absolute time, both theories postulate that $g_{\mu\nu}$ is a physical field defined on a background space-time. And, like Prokhovnik's view, the theory Pitts and Schieve consider entails that rulers and clocks are systematically distorted by the gravitational field in such a way that observers will conclude they inhabit a curved relativistic space-time \cite[1318]{PittsSchieve:2003}. In fact, Feynman (\citeyear{Feynman:2003}) and Michael Lockwood (\citeyear[335-336]{Lockwood:2007}) have utilized Poincare's creatures to explicate the theory.

There are multiple ways that a general relativistic metric can be laid on top of a Minkowski space-time. Pitts and Schieve argue that, for every point of the underlying Minkowski space-time, the gravitational field should have a well-defined value. In classical models of the Big Bang, there is no defined value for $g_{\mu\nu}$ prior to the cosmological singularity. So, to avoid postulating space-time points where the gravitational field is undefined, Pitts and Schieve lay $g_{\mu\nu}$ on top of the Minkowski space-time in such a way that the cosmological singularity is relegated to infinitely far in the past. There are no space-time points to the past of past time-like infinity, so there are no points of the underlying space-time where the gravitational field is undefined. If an analogous argument is applied to ANL, the cosmological singularity is again relegated to past time-like infinity. In that case, the universe would not have begun to exist at any time in the finite past.

Craig and Sinclair might reply that the gravitational field has a well-defined value at every point of the underlying space-time if the underlying space-time is truncated where the gravitational field becomes undefined; in that case, Craig and Sinclair would have reason to think that the underlying absolute time has a boundary. The trouble for this sort of view is two-fold. First, as I've discussed, there is a well-known and widely adopted principle according to which space-time should be maximally extended and that forbids the premature truncation of space-time. This was the principle that, in conjunction with relativistic cosmology, one might have hoped would help to overcome the Omphalos Objection in the first place. Second, Craig and Sinclair would still need a principled reason for choosing a specific metric for absolute time.

Pitts and Schieve argue that the only physically sensible way to overlay an FLRW metric as a gravitational field on a background Minkowski space-time banishes the cosmological singularity to the infinite past (at least as recorded by the metric of the underlying Minkowski space-time). Consequently, the cosmological singularity ``disappears'' \cite[1321]{PittsSchieve:2003}. For Craig and Sinclair to conclude that singular cosmological models depict the universe as having begun in the finite past according to absolute time, they must demonstrate that a similar verdict will not follow for ANL. Here, I turn to the third desideratum for establishing that the universe began to exist in the sense that Craig and Sinclair have defended, that is, that the duration of past absolute time over which the universe has existed is finite. In the next subsection, I develop a set of desiderata that Neo-Lorentzian accounts need to adequately satisfy in order to determine the direction and duration of time in a given cosmological model. I explicitly evaluate the classic big bang model in terms of each desideratum because, so far as I can tell, friends of ANL have not previously explicitly evaluated their project in terms of the desiderata. Moreover, I will re-use the results that I gather from evaluating the desiderata when I turn to bounce cosmologies in a later section. 

\subsection{Identifying the Span of Past Absolute Time}

In addition to the A-theory of time and a past boundary of the universe, in order to establish that the universe began, one must show that the span of absolute time since the past boundary to the present is finite. If one can only establish that the past universe has an open boundary,\footnote{Or some other pathology in virtue of which space-time is not further extendable to the past in whatever sense turns out to be appropriate.} then that open boundary may be located infinitely far into the past, in which case, at least in Craig and Sinclair's sense, the universe might not have begun to exist after all. That is, in addition to resolving the Omphalos Objection, one must resolve the Absolute Infinite Duration Objection, or ABIDO:

\begin{enumerate}
    \item If we do not know whether the absolute duration between the absolute present and the past boundary is infinite, then we do not know whether the universe began to exist in the finite past. 
    \item We do not know whether the absolute duration between the absolute present and the past boundary is infinite.
    \item Therefore, we do not know whether the universe began to exist in the finite past.
\end{enumerate}

There are four steps that friends of ANL should take to overcome ABIDO. First, friends of ANL should identify the requisite preferred foliation. Second, friends of ANL should determine a way to order the hypersurfaces in that foliation from the objective past to the objective future. Third, friends of ANL should identify a labeling of the hypersurfaces in the preferred foliation that corresponds to absolute time. And, fourth, friends of ANL should show that the total past duration of absolute time -- as measured by differences in the labeling of the hypersurfaces -- is finite.

Although Craig and Sinclair do not explicitly evaluate these four steps, I will show that the first two steps can plausibly be adequately addressed. However, I argue that Craig, Sinclair, and other friends of ANL who endorse a beginning of the universe in the finite past have not adequately addressed the third step. And since they have not adequately addressed step three, we will not be able to move on to step four. Without an adequately supported objective labeling of the hypersurfaces in their preferred foliation, friends of ANL cannot infer that the past had an objectively finite duration and so cannot infer that the universe began to exist a finite time ago.

\subsubsection{Step 1: Identify the preferred foliation}

Craig favors the view that space-time should be foliated into hypersurfaces of Constant Mean (extrinsic) Curvature (CMC). (For a non-technical introduction to the CMC foliation, see \cite[118-120]{Lockwood:2007}.) Consider a monotonically expanding FLRW space-time. Proper time, as recorded by observers who are co-moving with the universe's expansion, can be used to label the CMC hypersurfaces. This labeling is called the cosmic time. (However, the labeling is not unique; for example, the CMC hypersurfaces could instead be labeled with the scale factor.) The choice of the CMC foliation as the preferred foliation can be defended in several ways; collectively, they render plausible the choice of the CMC foliation as the preferred foliation for for friends of ANL. In passing, I note that the CMC foliation is unique only for closed universes. In the case of an open universe, there are an infinite collection of distinct foliations \cite[120]{Lockwood:2007}. Moreover, Michael Lockwood has argued that evidence for black hole decay is evidence that the actual universe has no CMC foliation \cite[152]{Lockwood:2007}. Here, I set these objections to one side in order to examine the case for accepting one of the CMC foliations as the preferred foliation.

Craig (\citeyear[236]{Craig:2001}) offers three arguments in support of his identification of a CMC foliation as the preferred foliation. First, Craig claims that the CMC foliation is ``natural'' because the foliation is defined by the global distribution of the universe's matter-energy content. Second, Craig draws upon an analogy with Newtonian spacetime. In Newtonian spacetime, the laws of motion assume a particularly simple form in inertial frames. For this reason, although we cannot identify which inertial frame corresponds to absolute space and time, one might argue that absolute space and time corresponds to one of the inertial frames. Likewise, in FLRW spacetimes, motion has a particularly simple form with respect to the CMC foliations and so one might surmise that one of the CMC foliations corresponds to absolute time.\footnote{An anonymous reviewer objects that the CMC foliation might not be the foliation picked out by the cosmic microwave background, as stipulated by Craig and other authors. As the reviewer notes, the CMB picks out a foliation for which the density of a scalar field -- representing the CMB -- is roughly spatially homogeneous, whereas a CMC foliation picks out a time-slicing so that the Hubble expansion is spatially homogeneous. Consequently, the two procedures could pick out distinct foliations. Supposing the two procedures did pick out distinct foliations, we would have yet another criticism of Craig's arguments and therefore further support to my own case against Craig's ANL. Nonetheless, I can see two reasons to think that the two procedures do pick out the same -- or approximately the same -- foliation.

First, we are discussing FLRW space-times, that is, space-times that are exactly homogenous and isotropic. Suppose that there were a space-time in which the CMB were not isotropic so that an observer co-moving with the universe's expansion would observe the CMB as being significantly ``hotter'' in one direction as compared with other directions. If the CMB were hotter in one direction than in another, then this would presumably be the result of an anisotropy in the matter-energy distribution. And if the matter-energy distribution is anisotropic, then the universe is not an FLRW space-time. So, while I agree that one could have had a CMB density that did not pick out a CMC foliation, I don't see how that would have been possible in an FLRW space-time. And given that the universe we inhabit is well approximated by the FLRW ansatz on cosmological scales, if everything else I've said in this paragraph is correct, we have that the CMB density picks out a foliation that is at least well approximated by the CMC foliation.

Second, in the case of FLRW space-times, space-time is ``naturally'' foliated in a way that locally corresponds to observers at rest with respect to the universe's expansion. This is the foliation that can be labeled with the cosmic time. And, as it turns out, at least in FLRW space-times, the surfaces of constant cosmic time are also surfaces of constant extrinsic scalar curvature. That is, the surfaces labeled by the cosmic time just are the CMC surfaces \cite[75]{Callender:2017}. But then the surfaces labeled by the cosmic time are just those that are uniform with respect to the CMB.} Third, in universes that approximate perfect homogeneity and isotropy on large scales, the Cosmic Background Radiation will appear, to a high degree of approximation, isotropic for the rest frames of CMC foliations. 

Simon Saunders has also argued that choosing a CMC foliation as the preferred foliation carries a number of theoretical advantages \cite[290]{Saunders:2002}. For example, the notion that one moment of time produces the next sits comfortably with the view that time objectively passes. Since CMC hypersurfaces are Cauchy surfaces, the full state of the world on one CMC hypersurface suffices for determining the state of the world on any other CMC hypersurface in the same foliation.\footnote{As Roser describes, 

\begin{quote}
    [...] the initial data can only be given on a a slice of constant scalar extrinsic curvature [that is, a CMC hypersurface], or equivalently of constant $T$. If we take the idea of a theory of gravity described by three-dimensional space whose geometry evolves through time (rather than the four-covariant `spacetime' picture) seriously, then the [fact that the initial data can only be given on a CMC hypersurface] strongly suggests that slices of constant $T$ are slices of constant time, so that the foliation on which the initial-value problem can be solved is indeed the foliation that corresponds to stacking of spaces at consecutive instances. For if physical time corresponded to a different time variable, that is, if the reconstruction of spacetime from the space-through-time theory were not a reconstruction from a constant-mean-curvature foliation, then as a consequence initial data could not be specified at a single instance in time. This would pose a major conundrum for the notion of what determines the dynamics of a physical system \cite[49]{Roser:2016}. 
\end{quote}
} Consequently, if the A-theorist's preferred foliation is one of the CMC foliations, then the A-theorist can imagine the state on one CMC hypersurface producing the state on a subsequent CMC hypersurface.

Craig has argued that effects which lie outside the domain of applicability of classical Special Relativity, e.g., quantum mechanics and cosmology, pick out a preferred foliation \cite[219-234]{Craig:2001}. Other friends of a preferred foliation agree. For example, Monton has argued that presentism may find a friendly home in quantum gravity approaches utilizing a fixed foliation into CMC hypersurfaces (\cite{Monton:2006}, though see the responses by W{\"u}thrich, i.e., \citeyear{Wuthrich:2010, Wuthrich:2013}). In addition, there are quantum gravity theories that violate Lorentz invariance by postulating a cut-off scale for the energy.\footnote{A cut-off scale does not necessarily imply Lorentz invariance violation; see, e.g.,  (\cite{RovelliSpeziale:2002vp}).} In turn, Lorentz invariance violation can result in a preferred reference frame, e.g.,  \cite[624-627]{Baron:2017}. Likewise, Ho\u{r}ava-Lifshitz gravity violates Lorentz invariance at high energy (\cite{Horava:2009, NilssonCzuchry:2019, TawfikDahab:2017}, \cite[132-134]{Koperski:2015}) and the CMC foliation has been suggested as the preferred foliation for Ho\u{r}ava-Lifshitz gravity (e.g., \cite{Afshordi:2009}). However, Ho\u{r}ava-Lifshitz gravity's relevance for Craig and Sinclair's project is unclear because some versions prevent singularities and lead to a bounce in the early universe (\cite{TawfikDahab:2017, Brandenberger2017}). Thomas Crisp (\citeyear{Crisp:2008, Crisp:2012}) and Pitts (\citeyear{Pitts:2004}) have both proposed presentist theories closely related to General Relativity and that utilize a CMC foliation. For example, Crisp has suggested that Julian Barbour's shape dynamics program can be adapted into a version of presentism that treats contemporary physics respectfully. However, Crisp and Pitts's proposals are incompatible with Craig and Sinclair's project, since Crisp and Pitts's proposals do not allow for an objective labeling of the CMC hypersurfaces that the theories pick out.

Some interpretations of quantum mechanics, e.g., Bohmian mechanics, violate Lorentz invariance, e.g., \cite[160-1]{Albert:1992}, \cite[59]{Valentini:1996}, and require a preferred foliation. Antony Valentini has suggested adopting a CMC foliation (and the York time for labeling hypersurfaces, as described below) in building a Bohmian cosmology \cite[60]{Valentini:1996}.

The point to take from this discussion is that if one is going to pick out a preferred foliation of our space-time as the one that corresponds to absolute time, then the CMC foliation is a particularly natural choice. Therefore, when Craig and Sinclair look for a preferred foliation of our space-time, they are right to pick out the CMC foliation as a suitable candidate.

\subsubsection{Step 2: Identify the ordering of the hypersurfaces\label{OrderOfTime}}

In order for Neo-Lorentzians to say that singular cosmologies (or, at any rate, cosmologies featuring space-times that are inextendable in whatever sense turns out to be appropriate) depict the universe as having begun to exist, they will need to show that our objective past has an open boundary represented by a singularity. And in order to show that our universe's objective past is bounded, they will need to provide an objective ordering of the hypersurfaces in the preferred foliation. If it should turn out that, according to absolute time, the cosmic singularity resides in our objective future, then the singularity provides no reason to think the absolute past is bounded. Some authors have thought that the direction of time shares a reductive explanation with the entropic arrow of time and so identify the direction of time with the direction of the entropy gradient. I will turn back to that view later, but, for now, note that authors who endorse Neo-Lorentzianism because of a prior commitment to A-theory, that is, most, perhaps all, friends of ANL, should not endorse the view that the direction of time has a reductive explanation.

Instead of utilizing the entropy gradient as an indication of the direction of time, Neo-Lorentzians can determine the direction of time from a relativistic description of chronogeometric structure. As Matthews (\citeyear{Matthews:1979}) and Castagnino (\citeyear{Castagnino:2003}) describe, a relativistic space-time -- which they denote $(M, g)$ -- admits of a global direction of time if $(M, g)$ satisfies three conditions (quoted from \cite[889--890]{Castagnino:2003}):

\begin{enumerate}
    \item $(M, g)$ is temporally orientable;
    \item For some $x \in M$, $(M, g)$ has a direction of time at $x$, that is, there is a non-arbitrary way of choosing the future lobe $C_x^+$ of the null cone $C_x$ at $x$;
    \item For all $x, y \in M$ such that $(M, g)$ has a direction of time at both $x$ and $y$, if the timelike vector $u$ lies inside $C_x^+$ and the timelike vector $v$ lies inside $C_y^+$ , then $u$ and $v$ have the same direction, that is, the vector resulting from parallel transport of $v$ to $x$ lies inside $C_x^+$.
\end{enumerate}

As shown in, e.g., \cite{Matthews:1979} and \cite{Castagnino:2003}, space-times satisfying the FLRW ansatz, i.e., those that are isotropic and homogoneous so that they can be described by the line element $ds^2 = -dt^2 + a(t)^2d\Sigma^2$, and that do not have pathological topological features (i.e., closed time-like curves) admit a global direction of time in this sense. However, that a space-time \emph{admits} of a global direction of time is merely to say that the space-time is compatible with a global direction of time. In order to identify the objectively correct ordering of the hypersurfaces in a given foliation, we would need a non-arbitrary method for choosing which of the two light cones at a given space-time point is the future (or past) cone. Unfortunately, relativity is inadequate, in itself, for determining which of the two cones is the future cone because the Einstein Field Equations are symmetric with respect to time. 

I offer two interrelated arguments for determining the direction of time from the relativistic chronogeometry that presuppose only that relativity is empirically adequate. First, the \emph{argument from empirical adequacy}. Note that for a theory to be \emph{empirically adequate}, the theory needs to make accurate predictions. Consequently, if relativity is empirically adequate within a given domain, all observations within that domain conform to the restrictions provided by the null-cone structure predicted from the Einstein Field Equations. On the A-theory, signals can be received only from the objective past and sent only to the objective future. Thus,

\begin{enumerate}
    \item If relativity is empirically adequate and A-theory is true, then any given local observer $o$ can receive signals only from points within $o$'s past light cone and transmit signals only to points within $o$'s future light cone.
\end{enumerate}

By definition, friends of ANL are committed to:

\begin{enumerate}[resume]
    \item Relativity is empirically adequate.
    \item A-theory is true.
\end{enumerate}

Friends of ANL can thereby conclude:

\begin{enumerate}[resume]
    \item Therefore, any given local observer $o$ can receive signals only from points within $o$'s past light cone and transmit signals only to points within $o$'s future light cone.
\end{enumerate}

But if $o$ can receive signals only from points within $o$'s past light cone and transmit signals only to points within $o$'s future light cone, then $o$ can use facts about their light cones to determine which light cone is objectively future and which is objectively past:

\begin{enumerate}[resume]
    \item If any given local observer $o$ can receive signals only from points within $o$'s past light cone and transmit signals only to points within $o$'s future light cone then $o$ can use facts about which points $o$ can receive signals from or send signals to to determine which of $o$'s light cones is objectively past and which is objectively future.
    \item Therefore, $o$ can use facts about which points $o$ can receive signals from or send signals to determine which of $o$'s light cones is objectively past and which is objectively future.
\end{enumerate}

Moreover, A-theorists think that we have immediate access to the passage of time. If we do have immediate access to the passage of time, then our immediate experience provides a non-arbitrary way of choosing the future lobe $C_x^+$ of the null cone $C_x$ at any given point $x$. And having picked out the future direction at $x$, we can parallel transport a future-directed time-like vector $u$ from $x$ to any other point $y$; if $u$ remains future-directed according to the locally defined direction of time at $y$, then the space-time has a global direction of time. All FLRW space-times, with the possible exception of those with sufficiently bizarre topological features, have a globally definable direction of time in this sense. Consequently, our phenomenology of time's passage provides us with an additional argument. First, let IMPAPT stand for ``immediate phenomenological access to the passage of time''. Then:

\begin{enumerate}
    \item If $o$ has IMPAPT then, for any local observer $o$, which of $o$'s two light cones is objectively past and which is objectively future can be determined from $o$'s IMPAPT.
    \item If which of $o$'s two light cones is objectively past and which is objectively future can be determined from $o$'s IMPAPT then which of the time-like tangent vectors along $o$'s worldline is future-directed can be determined using $o$'s IMPAPT.
    \item Therefore, if A-theory is true, which of the time-like tangent vectors along $o$'s worldline is future-directed can be determined using $o$'s IMPAPT.
\end{enumerate}

Call this the \emph{argument from IMPAPT}. According to friends of ANL, we have IMPAPT. Consequently, friends of ANL should be committed to the view that which of our time-like tangent vectors points into our future can be determined using our IMPAPT. Moreover, FLRW space-times admit of a global direction of time. Once a set of, e.g., future-directed time-like tangent vectors has been determined for $o$ at $p_1$, those vectors can be parallel transported to any other point $p_2$ in that space-time. As long as the space-time has a global direction of time in the sense defined above, the objective direction of time at $p_2$ is then the direction indicated by the parallel transported time-like tangent vector at $p_2$. Consequently, friends of ANL should be committed to the view that our IMPAPT allows us to determine a global direction of time.\footnote{An anonymous reviewer suggested a problem for the argument from IMPAPT. The argument from IMPAPT supposed that we could parallel transport, with respect to the Levi-Cevita connection, future directed tangent vectors from one space-time location to another. But, given the ANL proponent's interpretation of the Levi-Cevita connection, we might have reason to doubt the veridicality of results drawn from parallel transporting, with respect to the Levi-Cevita connection, vectors to arbitrary space-time points. To put the point another way, the argument from IMPAPT assumed that General Relativity is empirically adequate for a sufficiently broad class of potential (and not necessarily actual) observers. But if General Relativity is not empirically adequate for a sufficiently broad class of potential observers, the argument from IMPAPT would not establish the global direction of time. Similar worries may likewise endanger the cogency of the argument from empirical adequacy. Nonetheless, without a mathematically and empirically sufficient formulation of a Neo-Lorentzian successor to General Relativity, I have difficulty seeing how friends of ANL could establish the global direction of time in any other way. In order to be generous to Craig and Sinclair, I will assume that General Relativity is empirically adequate for a sufficiently broad class of potential observers, so that the size of the class of observers for whom General Relativity is empirically adequate is no problem for establishing a global direction of time using the arguments from empirical adequacy and IMPAPT.}

The arguments from empirical adequacy and from IMPAPT allow one to identify an objective ordering for the hypersurfaces in the CMC foliation of an FLRW space-time. I will return to these two arguments when I turn to bounce cosmologies below.

\subsubsection{Step 3: Identify a labeling of the preferred foliation}

I've argued that, given ANL, Craig and Sinclair may be able to defend the CMC foliation as the preferred foliation\footnote{Caveats apply since, for example, for FLRW space-times, the CMC foliation is unique only for closed universes.} and that that they can plausibly identify the objective ordering of the hypersurfaces in the preferred foliation. Nonetheless, as I show in this section, supposing that the difficulties in identifying a past boundary to the universe discussed in section \ref{IDing_the_boundary} can be overcome, I have difficulty seeing how Craig and Sinclair can adequately justify an objective measure of time from that past boundary to the present. For Neo-Lorentzians, each of the hypersurfaces in the preferred CMC foliation should be assigned a label that corresponds to the absolute time at which the events on that hypersurface take place. For Craig, the ``cosmic time plausibly coincides with God's metaphysical time, that is, with Newton's absolute time'' \cite[213]{Craig:2001}. And Craig has suggested that it is with respect to cosmic time that the universe can be said to have had an \emph{ex nihilo} beginning a finite time ago \cite[204]{Craig:2001}. Craig frequently moves back and forth between the CMC foliation and cosmic time as if the cosmic time labeling and the CMC foliation were equivalent.\footnote{For example, on page 220 of Craig's \citeyear{Craig:2001_Eternity}, Craig cites \cite{QadirWheeler:1985} in support of Craig's comments on cosmic time. While Qadir and Wheeler use the phrase `cosmic time' in their paper, Qadir and Wheeler's `cosmic time' is York time. Despite Qadir and Wheeler's placement of the Big Bang at past time-like infinity, Craig states -- on the same page! -- that cosmic time places the Big Bang at approximately fifteen billion years ago.} They are not. Foliations do not uniquely determine a labeling of the hypersurfaces that they pick out. Unfortunately, Craig's arguments for identifying the cosmic time as the objectively correct labeling of the preferred CMC hypersurfaces either do not uniquely pick out the cosmic time or else are not obviously stronger than arguments for other possible labelings.

Consider a space-time in which the line element is given by the FLRW ansatz, i.e., $ds^2 = -dt^2 + a(t)^2 d\Sigma^2$, where $t$ is the cosmic time, $a(t)$ is the scale factor, $d\Sigma^2$ is the spatial part of the line element (e.g., for flat space, $d\Sigma^2 = dx^2 + dy^2 + dz^2$). Suppose that $a(t)$ is a monotonic function, as is true for expanding or contracting universes. In that case, each of the CMC hypersurfaces in the preferred foliation can be labeled with $t$, $a(t)$, or with any bijective and order-preserving function of $t$, for example, inverse powers of $a(t)$, i.e., $\tau = -a(t)^{-n}$. (The mapping must be order-preserving in order to be consistent with the results of step 2.) Note that $\tau$ labels the cosmological singularity with $-\infty$ and the present with $\tau = 0$. Hans Halvorson and Helge Kragh argue that this raises doubts about whether the past finitude of the universe has ``intrinsic physical or theological significance'' (\cite{sep-cosmology-theology}). Several cosmologists have offered similar verdicts (\cite{Milne:1948, Misner:1969, Roser:2016, RoserValentini:2017}). On Charles Misner's view, the ``clock'' that we should use at ``early'' times is given by $\Omega = -\log(V^{1/3})$, where $V$ is the ``volume'' of the universe calculated as an inverse power of $a(t)$ \cite[1332]{Misner:1969}. Since the volume shrinks to zero as one approaches the cosmological singularity, $\Omega$ maps the cosmological singularity to infinitely far in the past. Misner concluded (emphasis his), ``\emph{The Universe is meaningfully infinitely old}'' \cite[186]{Misner:1969}.

Misner's cosmological model is outmoded. However, contemporary physicists have suggested labeling the CMC hypersurfaces with a quantity termed the \emph{York time} (\cite{York:1972}; \cite[49]{Roser:2016}). The York time is the trace of the extrinsic curvature $K$ of each CMC hypersurface. In the case of a model satisfying the FLRW ansatz, the York time is proportional to the negative Hubble parameter, i.e., $Tr K \propto -H$. Consequently, in FLRW models, there is an order preserving bijection between the York time and the cosmic time.

The point for our purposes is two-fold. First, as physicist Philipp Roser notes, ``even if there are other options [for a labeling that coincides with absolute time], York time must be considered a favourite among them based only on a few theoretical principles'' \cite[52]{Roser:2016}. Second, in FLRW space-times, York time varies inversely with $a(t)$ and, like $\tau$, the York time labels the cosmological singularity with negative infinity. As Roser describes, ``[...] just as in the case of Misner’s parameter [...] the `beginning' lies in the infinite past [...] The York-time approach to quantum gravity gives no explanation of a beginning because the universe simply has none. It is infinitely old'' (\cite[58--61]{Roser:2016}; also see \cite{RoserValentini:2017}). Importantly, if York time is at least as good a choice for absolute time as cosmic time, then we have reason to endorse the second premise in ABIDO, i.e., we cannot know whether the absolute duration between the present and the past boundary is infinite.

Craig does try to articulate some advantages of identifying absolute time with cosmic time. However, each of the supposed advantages of cosmic time that Craig recounts fail to uniquely pick out cosmic time. In part, this is because, as mentioned earlier, Craig does not consistently distinguish the preferred foliation from his labeling of the preferred foliation with cosmic time. For example, Craig tells us that observers whose clocks measure the cosmic time will record the CMB as isotropic. Nonetheless, the CMB is also isotropic according to, e.g., York time or with respect to any other labeling of the CMC hypersurfaces. One might think that cosmic time has the advantage that physically embodied observers whose clocks measure cosmic time are easier to come by than physically embodied observers whose clocks measure York time. True enough, but this cannot be an advantage for friends of absolute time. On their view, absolute time is independent of the time recorded by any local (or even physically constituted) clock. As Craig, channeling Newton, would tell us, local clocks need not record God's absolute time.

Perhaps, as an anonymous reviewer has objected, I've moved too fast and the fact that physically embodied observers possess clocks that record the cosmic time is an advantage for cosmic time. For example, if phenomenal conservatism is true, then, unless an argument to the contrary is available, we should assume that what seems to be true really is true. Local clocks seem to measure the time, so, unless we have an argument to the contrary, perhaps we should think that local clocks do measure the absolute time. And since local clocks approximate the cosmic time, we have reason to associate absolute time with cosmic time.

I am not persuaded by the phenomenal conservative argument. To start, note that although Craig thinks we have phenomenological access to the passage of time, there is a distinction between time's passage and the rate at which time passes. Local clocks do not actually measure cosmic time; instead, only the local clocks of observers who are co-moving with the universe's expansion approximate cosmic time. Supposing that absolute time should be identified with cosmic time, the rate of temporal passage experienced by any observer moving with respect to the universe's expansion is illusory. In fact, Craig argues that we are moving with respect to the universe's expansion \cite[56-57]{craig_2001}, so that Craig implicitly admits that, if cosmic time is absolute time, the rate at which we experience the passage of time is illusory. If the rate at which most observers experience the passage of time turns out to be a widespread illusion, then we have reason not to trust the rate at which time seems to pass.

Furthermore, the phenomenal conservative strategy suggests that, in our ordinary experience, clocks that can be physically constructed seem to measure time, so that perhaps we should think our local clocks measure absolute time. The suggestion followed that this gives us some reason to favor cosmic time as the measure of absolute time. However, we also have some seemings from attempts to construct quantum gravity theories or from attempts to construct an ontology for quantum theory that have elsewhere been argued to count in favor of York time as a candidate for absolute time. (Roser goes so far as to note that cosmic time is a ``highly unnatural choice of time parameter when discussing the very early universe'', \cite[58]{Roser:2016}.) Why should a metaphysics of absolute time favor the former seemings over the latter seemings? I don't see a clear way to settle the dispute between the two conflicting sets of seemings in favor of one set of seemings and that would be considered widely attractive to all disputants. And without a way to settle the dispute, all else being equal and provided absolute time exists, I am inclined to agnosticism concerning the finitude of the duration of past absolute time. And that agnosticism provides sufficient justification for endorsing ABIDO.

York time may have some advantages over cosmic time as a candidate for absolute time. Cosmic time can be used to label a CMC foliation of our universe provided our universe satisfies the FLRW ansatz, that is, if our universe is exactly homogeneous and isotropic. Our universe is not homogeneous and isotropic; instead, our universe only appears to be homogenous and isotropic when one averages the mass-energy content over sufficiently long spatio-temporal scales. As Gerald Whitrow writes, ``cosmic time is essentially a statistical concept, like the temperature of a gas'' (\citeyear[246]{Whitrow:1961}; also see \cite[117-118]{Lockwood:2007}). As Daniel Saudek notes, this leads to the consequence that cosmic time is defined only in a coarse grained way -- that is, for what Saudek calls ``stages'' of cosmological evolution -- and is inadequate for defining a total ordering over space-like separated localized events (\citeyear[56]{Saudek:2020}), as should be expected from a good candidate for absolute time. Thus, cosmic time can only be used as an approximate label of a CMC foliation of our universe. York time does not require homogeneity or isotropy; in fact, York time plausibly requires no averaging at all and so can be used as an exact label.\footnote{There may be worries about quantum indeterminacy. For example, perhaps there is no precise matter-energy distribution, with the consequence that the York time becomes ambiguous on small scales and so that no labeling could be exact. If so, we would be unable to define any candidate for the absolute time based on chronogeometric structure on sufficiently small spatio-temporal scales. Nonetheless, this would be an objection that applies equally to all possible candidates for absolute time and not to any specific candidate.}

Beyond our cosmological horizon, we have no reason to expect space-time to be homogenous or isotropic and so no reason for the FLRW ansatz to hold. For example, in an inflationary multiverse scenario, the distinct universes are proper parts of one space-time. Some -- though perhaps not all -- of the universes can be approximated using the FLRW ansatz so that \emph{a} cosmic time can be approximately defined for each universe.\footnote{George Ellis and Rituparno Goswami have promoted a generalization of the cosmic time -- the proper time co-moving gauge -- as a candidate for the absolute time (\citeyear[250]{EllisGoswami:2014}). Ellis and Goswami's proposal would apply to inhomogenous or anisotropic space-times. However, Ellis and Goswami's proposal labels a distinct foliation from the CMC foliation. Moreover, while one might have expected that any surface in the foliation labeled by absolute time is a space-like surface, Ellis's proposal has the bizarre consequence that, in inhomogenous space-times, some surfaces in the foliation Ellis's proposal picks out may be time-like surfaces. Therefore, if the proper time co-moving gaugage corresponds to absolute time, some moments of time are time-like surfaces, which is implausible. For this reason, the York time parametrization of the CMC foliation is arguably superior or at least not inferior.} However, the FLRW ansatz cannot be used to describe the \emph{entire} space-time, so that cosmic time cannot be defined as a global parameter for the entire multiverse. Nonetheless, supposing that the entire multiverse is a globally hyperbolic space-time manifold -- as is presumably required for ANL -- York time \emph{can} be defined as a global parameter.\footnote{Craig Callendar and Casey McCoy object that in the de Sitter phase of an inflationary multiverse, all CMC surfaces have the same value of the York time (\cite{CallendarMcCoy:2021}). If all of the CMC surfaces in the de Sitter phase have the same value of the York time, and the York time is identified with absolute time, then the awkward consequence follows that all of the CMC surfaces in the de Sitter phase are (somehow) \emph{simultaneous with} one another. Nonetheless, Roser points out that an actual inflationary phase is only approximately de Sitter and that the York time really is ``increasing during this cosmological period'' (\cite[50]{Roser:2016}; also see \cite[76-79]{Roser:2016}).}

Presentists maintain that only present objects exist simpliciter. There doesn't seem to be a coherent way for presentists to maintain that there are multiple presents, one present for each space-time region where the FLRW ansatz approximately holds. Therefore, insofar as we have either no reason to expect that space-time satisfies the FLRW ansatz beyond our cosmological horizon or that we even possess evidence that we do inhabit an inflationary multiverse (or some other anisotropic space-time) the presentist should favor a globally definable parameter as a candidate for absolute time. The York time is a globally definable parameter for a far broader class of space-times than cosmic time and is, for that reason, preferable by the presentist's own lights.

\section{Bounce Cosmologies}

I've issued two epistemological challenges to authors who endorse both ANL and the view that singular cosmological models support the conclusion that the universe had a beginning in the finite past. I now turn to considering how those who endorse Neo-Lorentzianism and the A-theory of time should interpret bounce cosmologies, that is, cosmological models according to which our expanding universe was birthed from a prior contracting universe. Bounce cosmologies can be constructed in classical General Relativity, but more sophisticated bounce cosmologies can be constructed in quantum gravity theories, such as Ho\u{r}ava-Lifshitz gravity (as already mentioned), loop quantum gravity (\cite{AshtekarSingh:2011, Cai:2014jla, AgulloSingh:2017}), string theory (\cite{IjjasSteinhardt:2017, Ijjas_2018, Ijjas:2019pyf, Steinhardt:2002, steinhardt_2007, ShtanovSahni:2003}), and $f(R)$ gravity (\cite{Corda:2011, Odintsov:2015, Oikonomou:2015}). All four result in modifications to the Einstein Field Equations and corresponding modifications to the FLRW equations. As discussed in \cite{Linford:2020b}, some bounce cosmologies depict our universe as having been produced from a black hole in another universe; in this paper, I've set aside cosmologies in which universes are born from ``bouncing'' through black holes, though analogous conclusions apply.

\subsection{The Thermodynamic Arrow of Time and Bounce Cosmologies}

Bounce cosmologies lack a singular boundary to time. For that reason, one might have thought that bounce cosmologies depict physical reality as not having a beginning, at least in Craig and Sinclair's sense. In order to defend the view that physical reality did have an absolute beginning, Craig and Sinclair set about trying to show that either bounce cosmologies are implausible or that, contrary to orthodoxy, bounce cosmologies depict physical reality as having an absolute beginning after all. Here, I set aside whether bounce cosmologies are empirically successful (or are otherwise plausible) in order to focus on the interpretation of bounce cosmologies. 

Following the previous section's discussion, let's consider a CMC foliation of a bounce cosmology. In many bounce cosmologies, the entropy obtains a minimum value at the CMC hypersurface (herein, the \emph{interface}) joining the two universes. On each of the hypersurfaces before the interface, the entropy decreases and, on each hypersurface after the interface, the entropy increases. In other contexts, the direction of increasing entropy has been termed the ``thermodynamic arrow of time'' because the direction in which entropy increases is (often) the future direction. Since the entropy increases in both directions away from the interface, Craig and Sinclair have argued that the future lies in both directions away from the interface. In other words, on their view, the interface is a closed boundary indicating the beginning of time for two universes. Craig and Sinclair write, ``The boundary that formerly represented the `bounce' will now [be interpreted to] bisect two symmetric, expanding universes on either side'' \cite[122]{CraigSinclair:2012}. Elsewhere, Craig and Sinclair state, ``The last gambit [in trying to avoid an absolute beginning], that of claiming that time reverses its arrow prior to the Big Bang, fails because the other side of the Big Bang is \emph{not} the past of our universe'' \cite[158]{CraigSinclair:2009}. And they conclude, ``Thus, the [universe on the other side of the interface] is not our past. This is just a case of a double Big Bang. Hence, the universe \emph{still} has an origin'' \cite[180-181]{CraigSinclair:2009}.

Nonetheless, as Daniel Linford (\citeyear{Linford:2020a, Linford:2020b}) has recently argued, there is tension between the views that the entropy gradient indicates the future direction and that the direction of time is irreducible. According to reductionists, such as David Albert (\citeyear{albert_2000, Albert:2015}) and Barry Loewer (\citeyear{Loewer_2007, Loewer:2012, Loewer:2019}), there is no microphysical distinction between past and future directions. Instead, the future appears to lie in a distinct direction from the past because the various macrophysical time asymmetries share a common reductive explanation with the entropy gradient. However, as Craig tells us, ``From the standpoint of a classical A-theorist like Isaac Newton, the failure of reductionistic attempts to explain the asymmetry of time is patent, since physical processes are at best mere sensible measures of time, not constitutive of time itself'' \cite[351]{Craig:1999}. Any attempt to reduce the direction of time is ``misconceived'' because God could have created a``universe lacking any of the typical thermodynamic, cosmological or other arrows of time'' but in which God ``experiences the successive states of the universe in accord with the lapse of His absolute time'' \cite[162]{craig_2001}. We don't need to adopt Craig's theology; as Poincar{\'e} wrote, ``the atheists [can imagine] put[ting] themselves in the place where God would be if he existed'' \cite[217]{Poincare:2001a}. One way that the universe could lack the typical arrows of time would be if the entropy decreased in the direction in which God experiences the successive states of the universe. We can imagine a God's eye view of the world in which eons pass as the entropy decreases and absolute time inexorably flows forward. Thus, Craig's metaphysics of time entails that the alignment between the direction of time and the thermodynamic arrow is not metaphysically necessary. In fact, Craig has argued that the ``physically reductionist accounts of temporal direction and/or anisotropy have little to commend themselves'' precisely because ``the physical arrows are neither necessary nor sufficient for time's having a direction and/or anisotropy'' \cite[352]{Craig:1999}.

Craig has suggested that the second law of thermodynamics may render the alignment between the direction of time and the entropic arrow nomologically necessary \cite[78]{CarrollCraig:2016}. Nonetheless, the second law of thermodynamics is a statistical regularity that admits of exceptions; violations of the second law of thermodynamics are not nomologically impossible. In some places, Craig appears to admit that there is no nomologically necessary connection between the entropy gradient and the direction of time.\footnote{For example,  Craig has considered a thought experiment in which the universe is a vast equilibrium gas with small, localized fluctuations from equilibrium. As Craig notes, for his reductionist interlocutors, there may be no sense in which a fluctuation at one time is either before or after a fluctuation at another distinct time. Craig thinks that an advantage of his anti-reductionism is that there would be a fact about which fluctuation is first regardless of how the universe's entropy changes in the interim \cite[354]{Craig:1999}. As Craig (\citeyear[355]{Craig:1999}) writes, ``The fact that entropy states of a process range in value between higher and lower numbers tells us nothing about which values exist later''.} Thus, on ANL, the alignment between the thermodynamic arrow and the direction of time is not logically, metaphysically, or nomologically necessary and Craig and Sinclair's interpretation of the interface as a double big bang cannot be adequately justified. The points that I've raised in this section are important and much more can be said. For now, let's move on by reminding ourselves of the resources that ANL supplies for determining the direction of time.

\subsection{The Global Direction of Time in Bounce Cosmologies}

We can imagine a CMC foliation of the bounce cosmology's space-time and a congruence of time-like curves passing through the foliation. And recall the procedures that we used in the Argument from IMPAPT and in the Argument from Empirical Adequacy. In both cases, we locally determined the direction of time -- either from our IMPAPT or from our ability to send or receive signals -- and then constructed the global direction of time by projecting the local direction via parallel transport. To determine whether or not time reverses at the interface, we can parallel transport future-directed time-like tangent vectors along a time-like geodesic congruence back through the interface and examine what the vectors do as they cross the interface. If we find that the future-directed, time-like tangent vectors become past-directed -- so that the direction of the objective past and the direction of the objective future flips at the interface -- then we would have reason to think that the interface is the beginning of two universes. And if the future-directed, time-like tangent vectors do not reverse their direction then the objective direction of time, as endorsed by Neo-Lorentzians, does not change at the interface. In that case, the other universe would be situated in our objective past and, contra Craig and Sinclair, bounce cosmologies should not be interpreted as having a beginning.

As I've said, bounce cosmologies can be constructed using either standard or modified FLRW equations. All standard and modified FLRW space-times, with the exception of those that allow some exotic topological features that are independently rejected by presentists (e.g., space-times featuring closed time-like curves), can be assigned a globally consistent direction of time because they satisfy the FLRW ansatz previously discussed. If we accept ANL's commitment to an absolute distinction between the past and the future in the current universe, then, not only does there exist a consistent time-ordering, but the time-ordering is uniquely determined by our IMPAPT and by the direction in which we can receive or transmit signals. In standard or modified FLRW space-times, the parallel transport of any future-directed time-like tangent vector along any time-like curve in any of the relevant FLRW space-times will never result in a past-directed time-like tangent vector. Therefore, when the direction of time is projected to the other universe via parallel transport, we find that the other universe is to our past. Therefore, contra Craig and Sinclair, friends of ANL should not interpret the interface as the birth of two universes.

One might object to the view that I've offered in this section on the grounds that, given the mismatch between the entropic arrow in the other universe, according to some bounce cosmologies, events in the other universe would happen in reverse order. One needs to be careful because, as \cite{Linford:2020b} points out, some bounce cosmologies maintain that the entropy per unit volume decreases as one approaches the bounce and entropy becomes hidden behind a cosmological horizon, even though the ``total'' entropy does not decrease. In that case, there may still be sensible thermodynamic development along a given time-like curve from the prior universe into the current universe. However, other bounce cosmologies predict anti-thermodynamic behavior as one approaches the bounce. Assuming the supervenience of mental states on brain states, observers might form memories in the reverse direction; if we could communicate with them -- and two-way communication does not seem likely -- they might tell us that they experience the passage of time in reverse. Perhaps we should listen to our trans-universal interlocutors and conclude, with Craig and Sinclair, that the other universe is not to our past after all.

Note that this argument is distinct from a closely related objection to some bounce cosmologies; according to the objection, we never observe a global reversal of the entropic arrow and this provides us good inductive reason to doubt bounce cosmologies in which there is a global reversal of the entropic arrow.\footnote{Nonetheless, as argued in \cite{Linford:2020b}, not all bounce cosmologies do include a global reversal of the entropic arrow of the sort that would be undermined by an inductive inference of this kind.} Whatever one might think of the merits of this objection, I am concerned with how friends of ANL should interpret bounce cosmologies and not with whether any bounce cosmology is plausible.

Consider a family of debunking arguments against the A-theory of time according to which the passage of time makes no difference to the physical world. Assuming mental states supervene on brain states, the same sequence of phenomenal states would be realized regardless of whether the A-theory of time is true \cite[14-15]{Price:1997}, \cite{Prosser:2000, Prosser:2007, Prosser:2013}.\footnote{Kristie Miller (\citeyear{Miller:2017}) has recently offered a related argument. Miller argues that, independent of the supervenience of mental states on brain states, all versions of A-theory on offer suggest that the passage of time can make no relevant difference to our phenomenal experience. If she is right, then our experience of temporal passage does not provide a reliable guide to the direction in which time passes.} On behalf of A-theorists, Sam Barron (\citeyear{Baron:2017}) argues that we have no way to rule out the possibility that the passage of time will make a difference to future physics and so (possibly) make a difference to brain states. The A-theorist can apply a corresponding view to bounce cosmologies; perhaps observers in the other universe would feel as though time passes in the direction towards the bounce, even if our current physics indicates that physical processes will happen in the ``reverse direction'' with respect to the bounce. If so, our trans-universal interlocutors would not say that their experience of time is reversed from ours. Perhaps other A-theorists would say that experiencing temporal passage in reverse is not metaphysically possible, though Craig seems unbothered by the possibility of observers unable to successfully detect the passage of absolute time \cite[354]{Craig:1999}.

In any case, thinking about how to respond to our trans-universal interlocutors has limited utility. We can compare (i) how friends of ANL should interpret a given cosmological model and (ii) what friends of ANL should say if the model turns out to correctly describe the world. If our world turns out to be correctly described by a bounce cosmology, and if there turn out to be creatures inhabiting the other universe who experience temporal passage in reverse, then the A-theorist would be left in a similar epistemic position as someone who discovers that some members of their world are brains-in-vats and who begin to wonder how they could know that they are not envatted themselves. Recall that, according to Craig, our IMPAPT provides an intrinsic defeater-defeater which ``overwhelms the objections brought against'' the ``objectivity of tense and the reality of temporal becoming'' \cite[138]{Craig:2000}. Perhaps Craig would likewise say that we have an intrinsic defeater-defeater for any objections brought against our access to the \emph{direction} of temporal becoming. If so, then, as with any agent who possesses an intrinsic defeater-defeater against global skepticism, observers who find themselves in a universe described by a bounce cosmology should retain their confidence that they have successfully identified the direction of time and so retain their confidence that the other universe is to our past.

In any case, in this paper, I am considering cosmological models ``from the outside'', that is, how friends of ANL should \emph{interpret} bounce cosmological models. Here, friends of ANL should conclude that, \emph{according to the model}, first, denizens of the prior universe may either experience anti-thermodynamic phenomena (if the supervenience of the mental on the physical fails) or else be subject to a widespread illusion about the direction of temporal passage and that, second, we, as denizens of the posterior universe, may either retain our confidence in the direction of temporal passage -- provided we have an intrinsic defeater-defeater for our access to the direction of temporal passage -- or else we would have to be much less confident about our experience of temporal passage than we can be in the actual world. And perhaps friends of ANL should be content that we currently have no evidence that we inhabit a bounce cosmology. Either way, we cannot conclude that the interface represents the \emph{ex nihilo} beginning of two universes.

\section{Conclusion}

Craig and Sinclair are not friends of metaphysical skepticism. They maintain thick metaphysical positions, \emph{viz}, the existence of God, a thick conception of the passage of time, a global present pervading the entire universe, and so on. Furthermore, they have frequently critiqued traditional interpretations of relativity on the grounds that those interpretations require positivist principles now largely considered \emph{pace} in anglophone philosophy of science. But despite their attempts to skirt metaphysical skepticism, Craig and Sinclair have endorsed positions that, as I have argued, invite a form of metaphysical skepticism antithetical to their larger project.

The beginning of the universe, as understood by Craig and Sinclair, cannot be directly observed. For that reason, empirical arguments for a beginning of the universe require us to conjoin observational data with a robust physical theory. In the case of singular (or appropriately inextendable) FLRW cosmologies, the conjunction of observational data and General Relativity may allow us to infer that the universe had an open boundary in the finite past, provided the difficulties discussed in section \ref{IDing_the_boundary} can be overcome. But, for Craig and Sinclair, an open boundary to space-time is insufficient for inferring that the universe began. When observational data is instead conjoined with Craig and Sinclair's Neo-Lorentzian alternative to relativity, we find ourselves unable to convincingly formulate the inferences that the universe has a past boundary or that the past boundary resides in the finite past.

Craig and Sinclair have also offered an alternative interpretation of bounce cosmologies on which space-time has a closed boundary in the finite past. As I've shown, careful examination of bounce cosmologies reveals that Craig and Sinclair's Neo-Lorentzianism leaves us without reason to infer that bounce cosmologies include a closed boundary in the finite past.

\section{Acknowledgements}

Thanks to Levi Greenwood, Philipp Roser, George Gale, Felipe Leon, Jeffrey Brower, Martin Curd, and Jacqueline Mariña's dissertation seminar for helpful discussions or feedback on this article.

\singlespacing

\bibliographystyle{plainnat}
\bibliography{references}

\end{document}